\documentclass[preprint,showpacs,showkeys,preprintnumbers,amsmath,amssymb]{revtex4}

\usepackage{graphicx}
\usepackage{dcolumn}
\usepackage{bm}

\newcommand{\cL}{{\cal L}}
\newcommand{\cLg}{{\cal L}_{g}}
\newcommand{\p}{\partial}

\begin{document}

\title{The total energy-momentum of the universe in teleparallel gravity}

\author{Yu-Xiao Liu$^1$}
\thanks{Corresponding author}
\email{liuyx@lzu.edu.cn}
\author{Zhen-Hua Zhao$^2$}
\email{zhaozhenhua@impcas.ac.cn}
\author{Jie Yang$^1$}
\author{Yi-Shi Duan$^1$}
\affiliation{$^1$Institute of Theoretical Physics, Lanzhou
University, Lanzhou 730000, P. R. China}
\affiliation{$^2$Institute of Modern Physics, Chinese Academy of
Sciences, Lanzhou 730000, P. R. China}

\begin{abstract}
We investigate the conservation law of energy-momentum in
teleparallel gravity by using general Noether theorem. The
energy-momentum current has also superpotential and is therefore
identically conserved. The total energy-momentum, which includes
the contributions of both matter and gravitational fields, is
given by the integral of scalar densities over a three-dimensional
spacelike hypersurface. As an example, the universe in
teleparallel gravity is investigated. It is shown that the total
energy-momentum vanishes independently of both the curvature
parameter and the three dimensionless coupling constants of
teleparallel gravity.
\end{abstract}



% 04.20.Cv Fundamental problems and general formalism
% 04.20.Fy Canonical formalism, Lagrangians,
%          and variational principles

%\date{\today}

\maketitle

\section{Introduction}\label{Sec1}

The definition of energy-momentum density for the gravitational
field is one of the most fundamental and controversial problems in
general relativity. However, some attempts to the problem leads to
prescriptions that are not true tensors. The first of such
attempts was made by Einstein who proposed an expression for the
energy-momentum distribution of the gravitational field. There
have been many attempts to resolve the energy problem in
Einstein's theory of General Relativity.

Teleparallel theories of gravity, which based on Weitzenb\"{o}ck
geometry \cite{Weitzenbock1923}, have been considered long time
ago in connection with attempts to define the energy of the
gravitational field \cite{Mol}. In the theory of the teleparallel
gravity, the curvature tensor vanishes identically and gravitation
is attributed to torsion \cite{Hayashi1978}. Furthermore, the
fundamental field in this theory is a nontrivial tetrad rather
than the metric. It is known that there exists no covariant,
nontrivial expression constructed out of the metric tensor.
However, covariant expressions that contain second order
derivatives of tetrad fields are feasible
\cite{BabakGrishchuk1999}.

The teleparallel equivalent of general relativity (TEGR)
\cite{Hehl,Kop,Muller,Nester,Maluf1,AndradePereira1997,Auccalla,PereiraPRD2006}
is an alternative geometrical description of Einstein's theory.
Recently, a method for dealing with the localization of the
gravitational energy had been presented in the Lagrangian
framework of the TEGR by Andrade, Guillen and Pereira
\cite{AndradeGuillen2000}. They botained an energy-momentum gauge
current for the gravitational field. The expression is a true
space-time and gauge tensor, can be reduced to M\o ller's
energy-momentum density of the gravitational field.

Subsequently, Blagojevi\'{c} and Vasili\'{c} investigated the
conservation laws associated with the asymptotic Poincar\'{e}
symmetry of spacetime in the general teleparallel theory of
gravity \cite{Blagojevic2001}. They obtained the improved form of
the canonical Poincarr\'{e} generators, which defines the
conserved charges of the theory. While Maluf and da Rocha-Neto
considered the Hamiltonian formulation of the teleparallel
equivalent of general relativity  \cite{Maluf2002}. The
gravitational energy-momentum is given by the integral of scalar
densities over a three-dimensional spacelike hypersurface.

In this paper, we would like to re-examine the energy-momentum
problem of teleparallel gravity with general Noether theorem. Our
purpose is to present the relationship between conservation
theorems and invariance properties of physical systems in
teleparallel theory. We will prove that the energy-momentum
current has also superpotential and is therefore identically
conserved.

The paper is arranged as follows. In Sec. \ref{Sec2}, a brief
review of teleparallel gravity is given. In Sec. \ref{Sec3}, we
give a general description of the scheme for establishing
covariant conservation laws in gravitational theory. In Sec.
\ref{Sec4}, we use the scheme to obtain a conservation law of
energy-momentum in teleparallel gravity. In Sec. \ref{Sec5}, we
calculate the total energy and momentum of the universe in
teleparallel gravity by superpotential. Sec. \ref{Sec6} is devoted
to some remarks and discussions.

\section{Review of teleparallel gravity}\label{Sec2}

Let us start by giving a simple review of the teleparallel gravity
theory (for the details, see Ref. \cite{Vargas0303034}). We use
the Greek alphabet $(\mu,\nu,\lambda, \cdots = 0, 1, 2, 3)$ to
denote indices related to spacetime, and the Latin alphabet $(a,
b, c, \cdots = 0, 1, 2, 3)$ to denote indices related to the
tangent space, assumed to be a Minkowski space with the metric
$\eta_{ab} = {\rm diag} (+1,-1,-1,-1)$. In the theory of the
teleparallel gravity, spacetime is represented by the
Weitzenb\"{o}ck manifold $W^{4}$, and the action is given by
\begin{equation}\label{ActionTele}
S =  \int d^{4}x \, h\, \left( \frac{1}{16 \pi G} S^{\lambda \mu
\nu} \; T_{\lambda \mu \nu}+ {\mathcal L_{M}}\right),
\end{equation}
where $h = \det(h^a{}_\mu$) with $h^a{}_\mu$ a tetrad field which
satisfies $g_{\mu \nu} = \eta_{a b} h^a{}_\mu h^b{}_\nu$,
$\mathcal L_{M}$ is the Lagrangian of the matter field, and
$S^{\lambda \mu \nu}$ is the tensor defined by the torsion
$T^{\lambda\mu\nu}$ of the Weitzenb\"ock connection $\Gamma
^{\lambda}{}_{\mu \nu}$
\begin{eqnarray}
S^{\lambda \mu \nu}  &=&c_1 T^{\lambda\mu\nu}
        +\frac{c_2}{2}\left(T^{\mu \lambda \nu}
        - T^{\nu \lambda\mu}\right), \nonumber \\
     && +\frac{c_3}{2}\left(g^{\lambda \nu} \;
            T^{\sigma\mu}{}_{\sigma}
            - g^{\lambda\mu} \;T^{\sigma \nu}{}_{\sigma}
           \right) \label{Slamunu} \\
T^{\lambda}{}_{\mu \nu} &=& \Gamma ^{\lambda}{}_{\nu \mu}
        -\Gamma ^{\lambda}{}_{\mu \nu}, \\
\Gamma^{\lambda}{}_{\mu\nu}&=&
        h_{a}{}^{\lambda}\p_{\nu}h^{a}{}_{\mu},
\end{eqnarray}
with $c_1$, $c_2$, and $c_3$ three dimensionless coupling
constants \cite{Hayashi1978}. It is well know that the
Weitzen\-b\"ock connection presents torsion but no curvature
\cite{Aldrovandi1995} and the curvature of the Weitzenb\"ock
connection vanishes identically as a consequence of absolute
parallelism. It is important to remark that, in this theory, the
fundamental field is a nontrivial tetrad rather than the metric.
For the specific choice
\begin{equation}
c_1=\frac{1}{4}, \quad c_2=\frac{1}{2}, \quad c_3=-1, \label{par}
\end{equation}
teleparallel gravity reduces to the so called teleparallel
equivalent of general relativity.

\section{Conservation laws in gravitational theory}\label{Sec3}

The conservation law is one of the important problems in
gravitational theory. It is due to the invariance of the action
corresponding to some transforms. In order to study the covariant
energy-momentum law of special systems, it is necessary to discuss
conservation laws by Noether theorem in the general case
\cite{Duan1963, Duan1987, Duan1988, Feng1999, Cho1995}. The action
of a system is
\begin{equation} \label{action}
I=\int_{\cal M} d^{4}x \cL(\phi^{A}, \p_{\mu}\phi^{A}),
\end{equation}
where $\phi^{A}$ are independent variables with general index $A$
and denote the general fields. If the action is invariant under
the infinitesimal transformations
\begin{eqnarray}
x^{\prime\mu} &=& x^{\mu}+\delta x^{\mu}, \label{transformation1}\\
\phi^{\prime A}(x^{\prime}) &=& \phi^{A}(x)
+\delta\phi^{A}(x),\label{transformation2}
\end{eqnarray}
and $\delta\phi^{A}$ vanishes on the boundary of $\cal M$, $\p
\cal M$, then following relation holds
\cite{Duan1963,Duan1988,Feng1995,LiuRSEnergy,LiuRSAngular}
\begin{equation} \label{NoetherTheorem}
\p_{\mu}(\cL\delta x^{\mu}+\frac{\p\cL}{\p\p_{\mu} \phi^{A}}
\delta_{0}\phi^{A} )+[\cL]_{\phi^{A}}\delta_{0}\phi^{A}=0,
\end{equation}
where
\begin{equation}
[\cL]_{\phi^{A}}=\frac{\p\cL}{\p\phi^{A}}-\p_{\mu}
\frac{\p\cL}{\p\p_{\mu}\phi^{A}},
\end{equation}
and $\delta_{0}\phi^{A}$ is the Lie variative of $\phi^{A}$
\begin{equation}
\delta_{0}\phi^{A}=\phi^{\prime A}(x)-\phi^{A}(x)
= \delta\phi^{A}(x)-\p_{\mu}
\phi^{A}\delta x^{\mu}.
\end{equation}

If $\cL$ is the total Lagrangian density of the system, the field
equation of $\phi^{A}$ is just $[\cL]_{\phi^{A}}=0$. Hence from
Eq. (\ref{NoetherTheorem}), we can obtain the conservation
equation corresponding to transformations (\ref{transformation1})
and (\ref{transformation2})
\begin{equation} \label{ConservationEq}
\p_{\mu}(\cL\delta x^{\mu}+\frac{\p\cL} {\p\p_{\mu}\phi^{A}}
\delta_{0} \phi^{A})=0.
\end{equation}
It is important to recognize that if $\cL$ is not the total
Lagrangian density, such as the gravitational part $\cL_{g}$, then
so long as the action of $\cL_{g}$ remains invariant under
transformations (\ref{transformation1}) and
(\ref{transformation2}), Eq. (\ref{NoetherTheorem}) is still valid
yet Eq. (\ref{ConservationEq}) is no longer admissible because of
$[\cL_{g}]_{\phi^{A}}\not=0$.

In gravitational theory with the tetrad as elementary fields, we
can separate $\phi^{A}$ as $\phi^{A}=(h_{a}{}^{\mu}, \psi^{B})$,
where $\psi^{B}$ is an arbitrary tensor under general coordinate
transformations. Suppose that $\cLg$ does not contain $\psi^{B}$,
then Eq. (\ref{NoetherTheorem}) reads
\begin{equation} \label{NoetherTheoremOfLg1}
\p_{\mu} \left( \cLg\delta x^{\mu}+\frac{\p\cLg}
{\p\p_{\mu}h_{a}{}^{\nu}} \delta_{0} h_{a}{}^{\nu} \right) +
[\cLg]_{h_{a}{}^{\nu}} \delta_{0}h_{a}{}^{\nu}=0.
\end{equation}
Under transformations (\ref{transformation1}) and
(\ref{transformation2}), the Lie variations are
\begin{equation} \label{LieVariationOfVierbein}
\delta_{0}h_{a}{}^{\mu} = h_{a}{}^{\nu}  \p_{\nu}
\delta x^{\mu} - \delta x^{\nu}
\p_{\nu}h_{a}{}^{\mu},
\end{equation}
Substituting Eq. (\ref{LieVariationOfVierbein}) into Eq.
(\ref{NoetherTheoremOfLg1}) gives
\begin{eqnarray}
\p_{\mu} \left[ \left( \cLg\delta^{\mu}_{\sigma}
-\frac{\p\cLg}{\p \p_{\mu}h_{a}{}^{\nu}}
\p_{\sigma} h_{a}{}^{\nu} \right) \delta x^{\sigma}
+ \frac{\p\cLg}{\p
\p_{\mu}h_{a}{}^{\nu}} h_{a}{}^{\sigma}
\p_{\sigma}\delta x^{\nu} \right] \nonumber
\end{eqnarray}
\begin{eqnarray} \label{NoetherTheoremOfLg2}
 +\;\; [\cLg]_{h_{a}{}^{\mu}}(h_{a}{}^{\nu} \p_{\nu}\delta x^{\mu} -
\delta x^{\nu} \p_{\nu}h_{a}{}^{\mu} )=0.
\end{eqnarray}
Comparing the coefficients of $\delta x^{\mu}, \delta
x^{\mu}_{,\nu} $ and $\delta x^{\mu}_{,\nu\lambda}$, we can obtain
an identity
\begin{equation}
[\cLg]_{h_{a}{}^{\nu}}\p_{\mu}h_{a}{}^{\nu} +
\p_{\nu}([\cLg]_{h_{a}{}^{\mu}}h_{a}{}^{\nu})=0.
\end{equation}
Then Eq. (\ref{NoetherTheoremOfLg2}) can be written as
\begin{eqnarray}
&&\p_{\mu} \left[ \left( \cLg\delta^{\mu}_{\sigma} -\frac{\p\cLg}{\p
 \p_{\mu}h_{a}{}^{\nu}} \p_{\sigma} h_{a}{}^{\nu} + [\cLg]_{h_{a}{}^{\sigma}}
 h_{a}{}^{\mu} \right) \delta
 x^{\sigma} \right. \nonumber \\
&&~~~~~~+ \left. \frac{\p\cLg}{\p \p_{\mu}h_{a}{}^{\nu}}
h_{a}{}^{\sigma} \p_{\sigma} \delta
 x^{\nu} \right]=0. \label{ConservationLaw1}
\end{eqnarray}
This is the general conservation law in the tetrad formalism of
spacetime. By definition, we introduce
\begin{eqnarray}
\tilde{I}^{\mu}_{\sigma}~&\equiv&\cLg\delta^{\mu}_{\sigma} -\frac{\p\cLg}{\p
\p_{\mu}h_{a}{}^{\nu}} \p_{\sigma} h_{a}{}^{\nu} + [\cLg]_{h_{a}{}^{\sigma}}
h_{a}{}^{\mu}, \label{Imusigma} \\
\tilde{V}^{\mu\sigma}_{\nu}&\equiv&\frac{\p\cLg}{\p \p_{\mu}h_{a}{}^{\nu}}
h_{a}{}^{\sigma}. \label{Vmusigmanu}
\end{eqnarray}
Then Eq. (\ref{ConservationLaw1}) gives
\begin{equation}\label{ConservationLaw2}
\p_{\mu}(\tilde{I}^{\mu}_{\sigma}
\delta x^{\sigma}+ \tilde{V}^{\mu\sigma}_{\nu} %
\p_{\sigma} \delta x^{\nu})=0.
\end{equation}
Eq. (\ref{ConservationLaw2}) is tenable under arbitrary
infinitesimal transformations, so we can compare the coefficients
of $\delta x^{\sigma}, \delta x^{\sigma}_{,\mu}$ and $\delta
x^{\sigma}_{,\mu\lambda}$ and obtain
\begin{eqnarray}
\p_{\mu}\tilde{I}^{\mu}_{\sigma}&=&0,\label{Result1}\\
\tilde{I}^{\mu}_{\sigma}~
&=&-\p_{\nu}\tilde{V}^{\nu\mu}_{\sigma},\label{Result2}\\
\tilde{V}^{\mu\sigma}_{\nu}
&=&-\tilde{V}^{\sigma\mu}_{\nu}.\label{Result3}
\end{eqnarray}
Eqs. (\ref{Result1})-(\ref{Result3}) are fundamental to the
establishing of conservation law in the theory of gravitation.

\section{Conservation law of energy-momentum in teleparallel gravity}\label{Sec4}

From Eqs. (\ref{ActionTele}) and (\ref{Slamunu}), we can rewrite
the gauge gravitational field Lagrangian as follows:
\begin{equation}\label{LgTmu}
\cLg=\frac{h}{16\pi G}\left(c_{1}
T^{\lambda\mu\nu}T_{\lambda\mu\nu}
+c_{2}T^{\mu\lambda\nu}T_{\lambda\mu\nu}
+c_{3}T^{\nu\mu}{}_{\nu}T^{\lambda}{}_{\mu\lambda} \right).
\end{equation}
The further expression of $\cLg$ is
\begin{equation}\label{LgTabc}
\cLg=\frac{h}{16\pi G}\left(c_{1}T^{abc}T_{abc}
+c_{2}T^{bac}T_{abc} +c_{3}T^{ab}{}_{a}T^{c}{}_{bc} \right),
\end{equation}
with $T_{abc}$ defined as
\begin{equation}
T_{abc}=h_{a\mu}(h_{c}{}^{\nu}\p_{\nu}h_{b}{}^{\mu}
-h_{b}{}^{\nu}\p_{\nu}h_{c}{}^{\mu}).
\end{equation}

For transformations $x^{\prime \mu}=x^{\mu}+h_{a}{}^{\mu}b^{a}$,
Eq. (\ref{ConservationLaw2}) implies
\begin{equation} \label{ConservationLaw3}
\p_{\mu}(\tilde{I}^{\mu}_{\sigma}h_{a}{}^{\sigma}
+\tilde{V}^{\mu\nu}_{\sigma} \p_{\nu} h_{a}{}^{\sigma})=0.
\end{equation}
From Einstein equations $h  T^{\mu}_{a}
=[\cL_{g}]_{h^{a}{}_{\mu}}$ and Eq. (\ref{Imusigma}), we can
express $\tilde{I}^{\mu}_{\nu}h_{a}{}^{\nu}$ as
\begin{equation} \label{Imua}
\tilde{I}^{\mu}_{\nu}h_{a}{}^{\nu}=\left(\cLg\delta^{\mu}_{\nu}
- \frac{\p\cLg}{\p
\p_{\mu} h^{a}{}_{\lambda} }\p_{\nu} h^{a}{}_{\lambda}
  \right) h_{a}{}^{\nu}+h
T^{\mu}_{a}.
\end{equation}
Defining
\begin{equation} \label{tmua}
h\; t^{\mu}_{a}=\left(\cLg\delta^{\mu}_{\nu}
- \frac{\p\cLg}{\p \p_{\mu} h^{a}{}_{\lambda}
}\p_{\nu} h^{a}{}_{\lambda}  \right) h_{a}{}^{\nu}
+ \frac{\p\cLg}{\p \p_{\mu}
h_{b}{}^{\nu} } h_{b}{}^{\sigma} \p_\sigma h_{a}{}^{\nu},
\end{equation}
and considering Eq. (\ref{Vmusigmanu}), we then have
\begin{equation} \label{IeVe}
\tilde{I}^{\mu}_{\sigma} h_{a}{}^{\sigma}
+ \tilde{V}^{\mu\nu}_{\sigma} \p_{\nu}
h_{a}{}^{\sigma} =h(T^{\mu}_{a}+t^{\mu}_{a}).
\end{equation}
So Eq. (\ref{ConservationLaw3}) can be written as
\begin{equation} \label{ConservationLawForm1}
\p_{\mu}[h (T^{\mu}_{a}+t^{\mu}_{a})]=0,
\end{equation}
or
\begin{equation}\label{ConservationLawForm2}
\nabla_{\mu}(T^{\mu}_{a}+t^{\mu}_{a})=0.
\end{equation}
This equation is the desired covariant conservation law of
energy-momentum in a teleparallel gravity system. $t^{\mu}_{a}$
defined in Eq. (\ref{tmua}) is the energy-momentum density of
gravity field, and $T^{\mu}_{a}$ to that of matter part. By virtue
of Eq. (\ref{Result2}), the expression on the LHS of Eq.
(\ref{IeVe}) can be expressed as divergence of superpotential
$V^{\mu\nu}_{a}$
\begin{equation} \label{Tt}
h (T^{\mu}_{a}+t^{\mu}_{a})=\p_{\nu}V^{\mu\nu}_{a},
\end{equation}
where
\begin{equation} \label{SuperV}
V^{\mu\nu}_{a}=\tilde{V}^{\mu\nu}_{\sigma}h_{a}{}^{\sigma}
=\frac{\p\cL_{g}}{\p \p_{\mu}
h_{b}{}^{\sigma}} h_{b}{}^{\nu} h_{a}{}^{\sigma}.
\end{equation}
Eq. (\ref{Tt}) shows that the total energy-momentum density of a
gravity system always can be expressed as divergence of
superpotential. The total energy-momentum is
\begin{equation}\label{Pa}
P_{a}=\int_{\Sigma} d\Sigma_{\mu} h(T^{\mu}_{a}+t^{\mu}_{a})
=\int_{S} dS_{\mu\nu}V^{\mu\nu}_{a},
\end{equation}
where $dS_{\mu\nu}
=\frac{1}{3!}\varepsilon_{\mu\nu\alpha\beta\gamma}
dx^{\alpha}\wedge dx^{\beta}\wedge dx^{\gamma}$. This conservation
law of energy-momentum in general relativity has the following
main properties: It is a covariant definition with respect to
general coordinate transformations. But the energy-momentum tensor
is not covariant under local Lorentz transformations, this is
reasonable because of the equivalence principle.

Now we calculate the expressions of $V^{\mu\nu}_{a}$ by using the
gravity Lagrangian density (\ref{LgTabc}) of teleparallel gravity.
The explicit expressions are
\begin{eqnarray} \label{VofTele}
V^{\mu\nu}_{a}&=&\frac{h}{8\pi G} \left[ ( h_{c}{}^{\mu} h^{\nu}_{b} %
- h_{c}{}^{\nu} h_{b}{}^{\mu} ) \left( c_{1}T_{a}{}^{bc} %
+ c_{2}T^{b}{}_{a}{}^{c}\right) \right. \nonumber \\
&&+ \left. c_{3}(h_{a}{}^{\mu} h_{b}{}^{\nu} - h_{a}{}^{\nu}
h_{b}{}^{\mu})T^{cb}{}_{c}\right].
\end{eqnarray}

\section{The energy-momentum of the universe in teleparallel gravity}\label{Sec5}

About two decades ago, Rosen \cite{Rosen1994} considered a closed
homogeneous isotropic universe described by the
Friedmann-Robertson-Walker (FRW) metric:
\begin{equation}
ds^{2}=dt^{2}-\frac{a(t)^{2}}{ \left( 1+r^{2}/4\right) ^{2}}\left(
dr^{2}+r^{2}d\theta ^{2}+r^{2}\sin ^{2}\theta \,d\varphi
^{2}\right) . \label{FRWeq2}
\end{equation}
Then using Einstein's prescription, he obtained the following
energy-momentum complex
\begin{equation}
\Theta_{0}^{\,0}=\frac{a}{8\pi }\left[ \frac{3}{\left( 1+r^{2}/4\right)
^{2}}-\frac{r^{2}}{\left( 1+r^{2}/4\right) ^{3}}\right] . \label{RosEq}
\end{equation}
By integrating the above over all space, one finds that the total
energy $E$ of the universe is zero. These interesting results
fascinated some general relativists, for instance, Johri {\em et
al.} \cite{Johri1995}, Banerjee and Sen \cite{BanSen1997} and Xulu
\cite{XuluIJTP2000}. Johri {\em et al.} \cite{Johri1995}, using
the Landau and Lifshitz energy-momentum complex, showed that the
total energy of an FRW spatially closed universe is zero at all
times irrespective of equations of state of the cosmic fluid. They
also showed that the total energy enclosed within any finite
volume of the spatially flat FRW universe is zero at all times.

Recently, Vargas \cite{Vargas0303034} considered the teleparallel
version of both Einstein and Landau-Lifshitz energy-momentum
complexes. His basic result is that the total energy vanishes
whatever be the pseudotensor used to describe the gravitational
energy. It is also independent of both the curvature parameter and
the three teleparallel dimensionless coupling constants. But he
worked with Cartesian coordinates, as other coordinates may lead
to non-physical values for pseudotensor, As remarked in Ref.
\cite{Rosen1993}.

In this section we calculate the total energy-momentum of the
homogeneous isotropic FRW universe by our conservation law in two
kinds of coordinates: sphere coordinates and Cartesian
coordinates.

\subsection{The energy-momentum in sphere coordinates}\label{Sec5_1}
The line element of the homogeneous isotropic FRW universe is given by
\begin{equation}\label{ds2 1}
ds^{2} = dt^{2} - \frac{a(t)^2}{(1+kr^2/4)^2}(dr^{2} + r^{2}d\theta^{2}+
r^{2}sin^{2}{\theta}d\phi^{2}),
\end{equation}
where $a(t)$ is the time-dependent cosmological scale factor, and
$k$ is the curvature parameter $k = 0, \pm 1$. The tetrad
components are
\begin{equation} \label{te1}
h^{a}{}_{\mu}= {\rm diag} \left(1,\;\frac{a(t)}{1+kr^2/4},
\;\frac{ra(t)}{1+kr^2/4},\;\frac{ra(t)\sin\theta}{1+kr^2/4}\right).
\end{equation}
Their inverses are
\begin{equation} \label{te2}
h_{a}{}^{\mu}= {\rm diag} \left(1,\;\frac{1+kr^2/4}{a(t)},
\;\frac{1+kr^2/4}{ra(t)},\;\frac{1+kr^2/4}{ra(t)\sin\theta}\right).
\end{equation}

From Eqs. (\ref{te1}) and (\ref{te2}), we can now construct the
Weitzenb\"ock torsion $T_{abc}$, whose nonvanishing components are
\begin{eqnarray}
T_{101}&=&T_{202}=T_{303}=-\frac{\dot{a}(t)}{a(t)},\nonumber \\
T_{212}&=&T_{313}=\frac{-4+kr^2}{4ra(t)},\\
T_{323}&=&-\frac{4+kr^2}{4ra(t)}\cot \theta, \nonumber
\end{eqnarray}
where a dot denotes a derivative with respect to the time $t$. For
$T_a=T^{b}_{ab}$, the calculated result is as follows
\begin{eqnarray}\label{Ta1}
T_{0}=3\frac{\dot{a}(t)}{a(t)},\;\;\; %
T_{1}=\frac{4-kr^2}{2ra(t)},\;\;\; %
T_{2}=\frac{4+kr^2}{4ra(t)}\cot\theta.
%\;\;\; T_{3}=0.  \nonumber
\end{eqnarray}
For superpotential $V^{\mu\nu}_{a}$, its non-zero components are
\begin{eqnarray}
 V^{01}_{0}&=&\frac{c_{3}(-4+kr^2)ra(t)\sin\theta}{\pi
  G(4+kr^2)^2},\;\;\;
  V^{02}_{2}=V^{01}_{1}/r,\nonumber \\  %\;\;\; %
 V^{02}_{0}&=&-\frac{c_{3}a(t)\cos\theta}{2\pi G(4+kr^2)},\;\;\; %
 V^{12}_{1}=-\frac{c_3 \cos\theta}{8\pi G},\nonumber \\
 V^{01}_{1}&=&-\frac{2(2c_1+c_2+3c_3)r^{2}\dot{a}a(t)
   \sin\theta}{\pi G(4+kr^2)^2}, \\
 V^{12}_{2}&=&-\frac{(2c_1+c_2+3c_3)(-4+kr^2)
    \sin\theta}{8{\pi} G(4+kr^2)},\nonumber \\
 V^{03}_{3}&=&V^{01}_{1}/(r\sin\theta),\;\;\;
  V^{13}_{3}=V^{12}_{2}/\sin\theta.\nonumber \\
 V^{23}_{3}&=&-\frac{(2c_1+c_2+3c_3)
  \cot\theta}{8{\pi} G r},\nonumber
\end{eqnarray}

Let us now calculate the total energy-momentum of the FRW universe
at the instant $x^0= t=$ constant, which is given by the integral
over the space section or at the infinite of the space. At the
region of the integral, i.e. $S=\p \Sigma=\Sigma|_{r\rightarrow
\infty}$, we have $dt=dr=0$, and Eq. (\ref{Pa}) becomes
\begin{equation}
P_{a}=\int_{S} dS_{01} V^{01}_{a}=\int_{S} d\theta d\varphi
V^{01}_{a}=\lim_{r\rightarrow \infty} \int^{\pi}_{0} d\theta
\int^{2\pi}_{0}d\varphi \;V^{01}_{a}. \label{PaSphere}
\end{equation}
Substituting above calculated results of $V^{01}_{a}$ into Eq.
(\ref{PaSphere}) yields
\begin{equation}
P_{a}=(0,0,0,0).\label{Pa1}
\end{equation}
So, the total energy and momentum of the closed (k=1), open
($k=-1$) and spatially flat ($k=0$) universes vanish.

\subsection{The energy-momentum in Cartesian coordinates}\label{Sec5_2}
Transforming from polar to Cartesian coordinates, the FRW line
element (\ref{ds2 1}) becomes
\begin{equation}\label{ds2 2}
ds^{2}= dt^{2}- \frac{a(t)^2}{(1+kr^2/4)^2}(dx^{2}+dy^{2}+dz^{2}).
\end{equation}
The tetrad components and their inverses are
\begin{equation} \label{eamu1B}
h^{a}{}_{\mu}= {\rm diag} \left(1,\;\frac{a(t)}{1+kr^2/4},
\;\frac{a(t)}{1+kr^2/4},\;\frac{a(t)}{1+kr^2/4}\right),
\end{equation}
\begin{equation} \label{eamu2B}
h_{a}{}^{\mu}= {\rm diag} \left(1,\;\frac{1+kr^2/4}{a(t)},
\;\frac{1+kr^2/4}{a(t)},\;\frac{1+kr^2/4}{a(t)}\right).
\end{equation}
From these above two equations, the nonvanishing components of
$T_{abc}$ and $T_a$ are constructed as follows:
\begin{eqnarray}
T_{101}&=&T_{202}=T_{303}=-\frac{\dot{a}(t)}{a(t)},~~
T_{221}=T_{331}=-\frac{kx}{2a(t)},\nonumber \\
T_{112}&=&T_{332}=-\frac{ky}{2a(t)},~~
T_{113}=T_{223}=-\frac{kz}{2a(t)},\\
T_{0}~~&=&3\frac{\dot{a}(t)}{a(t)},~~ %
T_{a}=-\frac{kx^{i}\delta_{ia}}{a(t)},~~ (a,i=1,2,3) \nonumber
\end{eqnarray}
For superpotential $V^{\mu\nu}_{a}$, its non-zero components are
\begin{eqnarray} \label{Vmunua}
V^{0i}_{0}&=&\frac{2c_{3}kax^{i}}{\pi G(4+kr^2)^2},\;\;\;(i=1,2,3) \nonumber \\
V^{0i}_{a}&=&-\frac{2(2c_{1}+c_{2}+3c_{3})\dot{a}a}{\pi G(4+kr^2)^2}\delta^{i}_{a},\;\;\;%
  (i,a=1,2,3)\nonumber \\
V^{21}_{2}&=&V^{31}_{3}=\frac{(2c_{1}+c_{2}+2c_{3})kx}{4\pi G(4+kr^2)}, \\
V^{12}_{1}&=&V^{32}_{3}=\frac{(2c_{1}+c_{2}+2c_{3})ky}{4\pi G(4+kr^2)},\nonumber \\
V^{13}_{1}&=&V^{23}_{2}=\frac{(2c_{1}+c_{2}+2c_{3})kz}{4\pi
G(4+kr^2)}.\nonumber
\end{eqnarray}

Now we can calculate the total energy-momentum. In Cartesian
coordinates, $dS_{0i}=x^{i}rd\Omega = x^{i}r\sin\theta d\theta
d\varphi$, so Eq. (\ref{Pa}) becomes
\begin{eqnarray}\label{PaCartesian}
P_{a}&=&\int_{S} dS_{0i} V^{0i}_{a}=\int_{S}d{\theta}d{\varphi} \;
        x^{i}~r\sin\theta \; V^{0i}_{a} \nonumber \\
     &=&\lim_{r\rightarrow \infty} \int^{\pi}_{0} d\theta
        \int^{2\pi}_{0}d\varphi \;
        \left(x^{i}~r\sin\theta \; V^{0i}_{a}\right).
\end{eqnarray}
Substituting above calculated results into Eq. (\ref{PaCartesian})
yields again
\begin{equation}\label{Pa2}
P_{a}=(0,0,0,0).
\end{equation}
So, the values of the total energy and momentum of the FRW
universe, which calculated in Cartesian coordinate, vanish too.

\section{Summary and discussions}\label{Sec6}
To summarize, by the use of general Noether theorem, we have
obtained the conservation law of energy-momentum in teleparallel
gravity theory. The energy-momentum current has also
superpotential and is therefore identically conserved. Based on
this conservation law of energy-momentum, we have calculated the
total energy and momentum of the FRW universe, which includes
contributions of matter and gravitational field. All calculated
values vanish. They are also independent of both the curvature
parameter and the three teleparallel dimensionless coupling
constants. Therefore it is valid not only in the telaparallel
equivalent of general relativity, but also in any teleparallel
model. Commonly, the universe is filled with matter field and
gravitational field, but its total energy is actually zero. Thus
we can conclude that the gravitational energy exactly cancels out
the matter energy.

It is important to remark that, all results, calculated in both
sphere and Cartesian coordinates, are same. So the corresponding
conservative quantities are independent of the choice of
coordinates, which are caused by the covariance of the
conservation law. We think a covariant conservation law of angular
momentum is still needed in order to understand the conservative
quantities in the theory of teleparallel gravity.
% This conservation law can be obtained using the approach in
% \cite{Duan1988,Duan1995}.

\section*{Acknowledgement}
One of the authors (Yu-Xiao Liu) would like to express his
gratitude to Prof. Jian-Xin Lu and Prof. Rong-Gen Cai for their
suggestive discussions and hospitality. He also thanks Li-Ming
Cao, Li-Jie Zhang, Miao Tian, Zhen-Bin Cao and Li-Da Zhang for
helpful discussions. This work was supported by the National
Natural Science Foundation and the Fundamental Research Fund for
Physics and Mathematic of Lanzhou University (No. Lzu07002)

\end{document}